\documentclass[twocolumn,pra,tighten,floatfix,showpacs]{revtex4}

\usepackage{graphicx}% Include figure files
\usepackage{latexsym}
\def\<{\langle}
\def\>{\rangle}

\def\half{{\frac{1}{2}}}
\def\be{\begin{equation}}
\def\ee{\end{equation}}
\def\bea{\begin{eqnarray}}
\def\eea{\end{eqnarray}}

\def\ceff{c_{\rm{eff}}}

\begin{document}
%\preprint{cond-mat} \title{Nonlocal fermions and extensive entropy}
\preprint{cond-mat} \title{Nonlocal fermions and the entropy volume law}

\author{G. C. Levine}

\address{Department of Physics and Astronomy, Hofstra University,
Hempstead, NY 11549}
%\address{$^\dagger$Department of Physics, Sam Houston State University, Huntsville TX 77341}

\date{\today}

\begin{abstract}
To produce a fermionic model exhibiting an entanglement entropy volume law, we propose a particular version of nonlocality in which the energy-momentum dispersion relation is effectively randomized at the shortest length scales while preserving translation invariance. In contrast to the ground state of local fermions, exhibiting an entanglement entropy area law with logarithmic corrections, the entropy of nonlocal fermions is extensive, scaling as the volume of the subregion and crossing over to the anomalous fermion area law at scales larger than the locality scale, $\alpha$. In the 1-d case, we are able to show that the central charge appearing in the universal entropy expressions for large subregions is simply related to the locality scale. These results are demonstrated by exact diagonalizations of the corresponding discrete lattice fermion models. Within the Ryu-Takayanagi holographic picture, the relation between the central charge and the locality scale suggest a dual spacetime in which the size of the flat UV portion and the radius of AdS in the IR are both proportional to the locality scale, $\alpha$. 
\end{abstract}

%Several lattice models of spinless fermions featuring nonlocal kinetic energy and translation invariance are constructed. This volume entropy law is discussed in connection with proposals of quantum field theory models dual to flat space gravity. 

\maketitle
\section{Introduction} 

%Nonlocal field theories have been proposed as a quantum field theory model that may be dual to flat gravitational spacetime \cite{Li:2010dr,Shiba:2013jja,Bagchi:2014iea,Pang:2014tpa}. Specifically, it has been shown that the von Neumann entanglement entropy obeys a volume rather than area law. 

Nonlocal discrete lattice models and field theories have attracted attention recently in several---possibly related---subfields of physics, spanning quantum information, condensed matter and high energy physics. Random spin models such as \cite{Sachdev:2010um} have been introduced in connection with the phenomenology of strange metals, but also appear to be important in constructing exactly solvable quantum models with gravitational duals \cite{Maldacena:2016hyu}. Nonlocal models have also appeared in the study of the thermalization hypothesis providing the rigorous basis of the canonical ensemble of statistical mechanics \cite{Magan:2015yoa}.  Attempts to understand the internal degrees of freedom of black holes, consistent with the no-cloning theorem, have led to the notion of "fast-scrambling" \cite{Hayden:2007cs,Sekino:2008he} which also can be realized with certain nonlocal models \cite{Magan:2016ojb,Swingle:2016var}.  So far the fermionic and spin models considered either explicitly break translation invariance or simply do not have spatial degrees of freedom in the conventional sense.
 
This work attempts to extend nonlocality to fermionic models while maintaining translation invariance. The Maximally Entangled Renormalization Ansatz (MERA) provides a basis for the construction of hamiltonians with a prescribed long distance entanglement law. Motivated by the continuous version of MERA \cite{Haegeman:2011uy}, Li and Takayanagi \cite{Li:2010dr,Shiba:2013jja} proposed a nonlocal action of the form
\be
\label{NL_bosons}
S = \int{d^dx \phi(x) e^{-\alpha \sqrt{\partial^\mu \partial_\mu}} \phi(x)}
\ee 
where $\alpha$ may be thought of as the scale of nonlocality.  %Backed up by numerical calculations on lattice. %

However, the extension to fermions poses a difficulty seen in the following free (non-relativistic) fermion hamiltonian:%nonlocal chiral%
\be
\label{bad_NL_fermions}
H = \epsilon\int{dx \psi^\dagger(x) e^{-\half\alpha^2 \partial^2_x} \psi(x)}
\ee
Considering the corresponding discrete model on a periodic lattice of $R$ sites, $H = \sum_{xy}{c^\dagger_x e^{S_{xy}} c_y}$ (see eq. (\ref{LatticeDerivatives})), the nonlocal energy dispersion is $\epsilon_n = \epsilon e^{-\alpha^2 \cos{k_n}}$ where the wavenumber $k_n = 2 n \pi/R$ and $n, x \in [-\frac{R}{2},\frac{R}{2}]$. The zero temperature correlation functions of the half occupied band, $G(x) = \< c^\dagger_x c_0\>$ may be written:
\begin{equation}
\label{NL_L_GF}
G(x) = \frac{1}{R} \sum_{n=-R/2}^{R/2}{\theta(\epsilon_{\rm F} - \epsilon_n) e^{i k_n x}} =  \frac{1}{R} \sum_{n=-R/2}^{R/2}{\theta(-k_n)e^{i k_n x}}
\end{equation}
where $\theta$ is the unit step function and the Fermi energy, $\epsilon_{\rm F} = \epsilon$. 

Since the nonlocal energy dispersion function preserves wavenumber order of the eigenvalues that is identical to the conventional dispersion ($\epsilon_n = -\cos{k_n}$), the correlation functions in local and nonlocal cases (equations (\ref{NL_L_GF})) are also identical at zero temperature.  Entanglement entropy may be computed by a a standard procedure \cite{Peschel2003} from lattice correlation functions for noninteracting fermions, thus nonlocality in the hamiltonian above is not expected to produce a different entropy area law for fermions. In contrast to the last relation in equation (\ref{NL_L_GF}), boson correlation functions explicitly involve the dispersion relation and therefore reflect the nonlocality of the kinetic energy in (\ref{NL_bosons}).  

To overcome this difficulty, we propose models of the following type:
\be
\label{NL_fermions}
H =  \epsilon\int{dx \psi^\dagger(x) \cos{(i\alpha \partial_x)} \psi(x)}
\ee
As in (\ref{NL_bosons}) and (\ref{bad_NL_fermions}) the kinetic energy operator involves derivatives of all orders and nonlocality.  However, If $\alpha/\pi$ is large and irrational, nearby energy states below the Fermi level will be drawn from widely disparate wavenumbers---a feature we refer to as {\sl compact nonlocality}: maintaining the energy volume of the Fermi sea but scrambling the energy order of the translation invariant eigenfunctions.  

This model has two interesting features. First, $\alpha \rightarrow 0$ reproduces nonrelativistic fermions with a quadratic dispersion. Second, in the opposite limit $\alpha \rightarrow \infty$, the irrational values of $\alpha/\pi$ lead to random correlation functions resembling those of models with quenched disorder, although the present hamiltonian is translation invariance. This is the specific feature leading to the volume entropy law.  

Quantum many body states exhibiting an entropy volume law were first realized as excited states of integrable hamiltonians \cite{Alba:2009th} and some interesting aspects deeply explained in \cite{2015PhRvB..91h1110L} (see also \cite{Vidmar:2018rqk}). Viewed this way, the present work may bear some relation to this earlier work in that the ground state of our model for a given locality parameter $\alpha$ must correspond to a particular excited state of the corresponding local model. Recently there has also been interesting work on constructing exotic spin models exhibiting an entropy volume law in their ground state \cite{Klich2016a,Salberger:2016noc}. As we only consider realizations of the hamiltonian (\ref{NL_fermions}) on a lattice, this work represents only one possible approach to a constructing a theory with a volume law for fermions. Much analytical work has been accomplished on nonlocal QFT and flat space holography (\cite{Bagchi:2014iea,Kachru:2018,Pang:2014tpa}) and it is quite possible that there are other translationally invariant approaches to nonlocal fermions.

\section{nonlocal fermionic models}

Nonlocal bosons defined by (\ref{NL_bosons}) have been shown analytically \cite{Li:2010dr} and numerically \cite{Shiba:2013jja} to possess a volume entropy law.  As mentioned in the introduction, the difference between boson and fermion behaviors may be traced to the presence of the dispersion relation in the correlator, a feature of relativistic bosons.  Real space field operators have particle and antiparticle contributions: $\phi(x) \sim \sum{\sqrt{\frac{1}{\omega_k}}(a_k + a_{-k}^\dagger)e^{i k x}}$ where the Bogolubov coefficients explicitly involve the energy dispersion.  Fermionic superconductors share this feature with relativistic bosons, having real space field operators that mix particles and antiparticle components and a relativistic dispersion relation with the superconducting gap playing the (momentum dependent) role of particle mass. To this end it is natural to try to construct a model of nonlocal fermions by applying nonlocal extensions to fermionic superconductors. We will show, however, that the problems associated with nonlocal fermions described in the introduction persist even in the relativistic (superconducting) model and necessitate compact nonlocal fermionic extensions. To set the notation used in this manuscript it is convenient to start with the superconducting model before discussing models akin to (\ref{NL_fermions}).

% fix 2 pi n/L ---> 2 pi n/R

Consider a one-dimensional pairing hamiltonian for spinless fermions:
\begin{equation}
\label{ham}
H = \sum^{R-1}_{x,y=0}{(c^{\dagger}_{x}A_{xy}c_{y}+ \frac{1}{2}[ c^{\dagger}_{x}B_{xy}c^\dagger_{y} -c_{x}B_{xy}c_{y} ] )}
\end{equation}
where $x$ and $y$ are one-dimensional site indices and $A_{xy}$ and $B_{xy}$ are real kinetic energy and pairing matrices respectively, obeying  $A_{xy} = A_{yx} $ and $B_{xy} = -B_{yx}$. the operators $c_{x}$ ($c^{\dagger}_x,$) destroys (creates) fermions at site $x$ of a periodic $R$ site lattice and obey the conventional fermion algebra. Applying a Bogoliubov transformation 
\bea
\eta_n &=& \sum^{R-1}_{x=0}{(u_{nx} c_x + v_{nx} c_x^\dagger)} \\
\eta^\dagger_n &=& \sum^{R-1}_{x=0}{(u_{nx} c^\dagger_x + v_{nx} c_x)} 
\eea
brings (\ref{ham}) into diagonal form:
\be
H = \sum_n{E_n \eta_n^\dagger \eta_n}
\ee
where $\{u_{nx} \}$ and $\{v_{nx} \}$ are real coefficients, satisfying a lattice form of the Bogoliubov-de Gennes equation:
\be
\label{BdG}
\left(\begin{array}{cc} \mathbf{A} & \mathbf{B} \\ -\mathbf{B} & -\mathbf{A} \end{array}\right) \left(\begin{array}{c} \vec{u}_n \\ \vec{v}_n \end{array}\right) = E_n \left(\begin{array}{c} \vec{u}_n \\ \vec{v}_n \end{array}\right)
\ee
 The notation in (\ref{BdG}) has been simplified by writing the coefficients $\{u_{nx} \}$ and $\{v_{nx} \}$ as real space vectors, referenced by energy index $n$.
 Defining symmetric and antisymmetric lattice derivatives:
\begin{eqnarray}
\label{LatticeDerivatives}
S_{xy} &=& \half(\delta_{x,y+1} + \delta_{x,y-1}) \\
T_{xy} &=& \half(\delta_{x,y+1} - \delta_{x,y-1}) \nonumber
\end{eqnarray}
The conventional dispersion relation of superconducting quasiparticles is obtained by choosing nearest neighbor kinetic and pairing matrices, $A_{xy} = S_{xy}$ and $B_{xy} = T_{xy}$:
\be
E_n^2 = t_n^2 + \Delta^2_n
\ee
where $t_n \equiv \cos{(2 \pi n/R)}$ and $\Delta_n \equiv \sin{(2 \pi n/R)}$.  We will refer to this as the local relativistic dispersion relation. Although other choices will be discussed at the end of this section we presently choose matrices for $\mathbf{A}$ and $\mathbf{B}$ that are symmetric and antisymmetric, respectively, and have compact nonlocality features.
\bea
\label{NL_AB}
A_{xy} &=& [\cos{(\alpha \mathbf{S})}]_{xy} = [\cos{\tilde{\partial}}]_{xy} \\
B_{xy} &=& [\sinh{(-\alpha \mathbf{T})}]_{xy} =-i [\sin{\partial}]_{xy}  \nonumber
\eea
With the definitions $\tilde{\partial} \equiv \alpha \mathbf{S}$ and $\partial = i \alpha \mathbf{T}$, these matrices satisfy $\partial^\dagger = \partial$ and $\tilde{\partial}^\dagger = \tilde{\partial}$, and the nonlocal Bogoliubov de Gennes equation may written compactly:

\be
\label{NL_BdG}
\left(\begin{array}{cc} \cos{\tilde{\partial}} & -i\sin{\partial} \\ i\sin{\partial} & -\cos{\tilde{\partial}} \end{array}\right) \left(\begin{array}{c} \vec{u}_n \\ \vec{v}_n \end{array}\right) = E_n \left(\begin{array}{c} \vec{u}_n \\ \vec{v}_n \end{array}\right)
\ee
or
\bea
H\vec{\phi}_n &=& E_n \vec{\phi}_n \\
H = \sigma_z \cos{\tilde{\partial}} &+& \sigma_y \sin{\partial}
\eea
The energy dispersion now exhibits compact nonlocality in both kinetic and gap (mass) components:
\be
E_n^2 =  \cos^2{( \alpha t_n )} + \sin^2{( \alpha \Delta_n )} 
\ee

We now turn to models without pairing, and in particular, those that reduce to nonrelativistic lattice fermions in the local limit $\alpha \rightarrow 0$.  As mentioned in the Introduction, eq. (\ref{NL_fermions}) reduces to nonrelativistic fermions with quadratic dispersion as $\alpha \rightarrow 0$.  with $\epsilon= 1/\alpha^2$ identified as the nonrelativistic localization energy scale of a mass $m$ particle: $\hbar^2/2m\alpha^2$ (and $\hbar^2/2m = 1$).  Using eq. (\ref{LatticeDerivatives}),  the lattice form of equation (\ref{NL_fermions}) may be written:
\be
\label{NL_cos}
H = \epsilon\sum_{xy} {c^\dagger_x [\cos{\partial}]_{xy} c_y}  
\ee
Extending the definition of the symmetric lattice derivative, $S_{xy}$, to higher dimensions, the following hamiltonian has a quadratic dispersion local limit ($\alpha \rightarrow 0$) in any spatial dimension:
\be
\label{NL_sin}
H = \epsilon\sum_{xy} {c^\dagger_x [\sin{\tilde{\partial}}]_{xy} c_y}  
\ee
In the next section entanglement entropy for variations of these models will be explored and compared.

%Chosen for maintaining features of the band structure in the local superconducting counterpart.
\section{numerical results}

Following the procedure developed by Peschel \cite{Peschel2003} we compute the von Neumann entropy for various nonlocal, quadratic fermion models based upon hamiltonians (\ref{ham}, \ref{NL_AB}), (\ref{NL_cos}) and (\ref{NL_sin}).  For computing the entropy, the relevant greens functions are 
\bea
G_{xy} &\equiv& \<c^\dagger_x c_y \> = \sum_{n} {v_{nx} v_{ny}} \\
F_{xy} &\equiv& \<c^\dagger_x c^\dagger_y \> = \sum_{n} {v_{nx} u_{ny}} 
\eea
where the coefficients $\{ u_{nx} \}$ and $\{ v_{nx} \}$ were found by diagonalizing the BdG hamiltonian (\ref{BdG}). %(details GF GF pair B coefficients, stability of analytical/numerical approaches here). %
The entanglement entropy of a subregion is then computed by extracting the $\mathbf{G}$ and $\mathbf{F}$ matrices for the subregion and computing the entanglement eigenvalues $\{ \epsilon_l \}$ from 
\bea
\sum_{xy} (G_{zx} - F_{zx} &-&\half \delta_{zx}) ( G_{zx} + F_{zx} -\half \delta_{zx}) P_{ly} \nonumber \\ 
&=& \frac{1}{4} \tanh^2{\frac{\epsilon_l}{2}} P_{lz}          
\eea
Finally, the entropy is given by:
\be
S = \sum_l{ (\log(1+e^{-\epsilon_l}) + \frac{\epsilon_l}{1+e^{\epsilon_l}} ) }
\ee

\begin{figure}[ht]
\includegraphics[width=7.5cm]{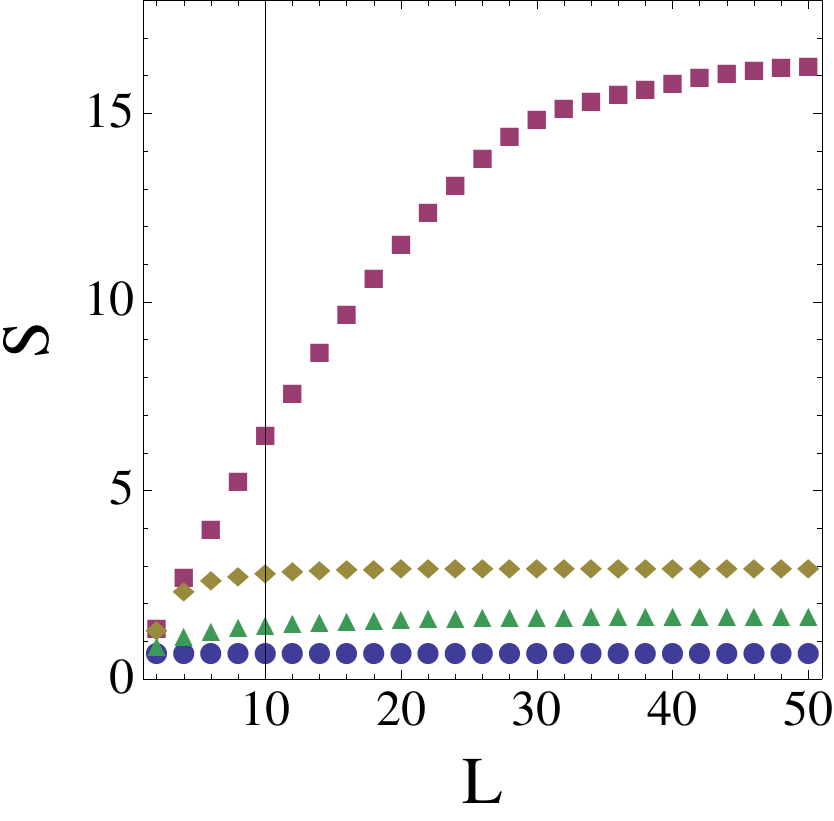}
\caption{\label{fig1} Entropy ($S$) versus length ($L$) of a subregion for several local and nonlocal models on a one-dimensional, 100-site lattice. Entropy computed for local, gapped superconducting hamiltonian (equation \ref{ham}) with kinetic and pairing matrices given by $\mathbf{A} = \mathbf{S}$ and $\mathbf{B} = \mathbf{T}$ (equations \ref{LatticeDerivatives}) ($\circ$). Entropy computed for noncompact, nonlocal hamiltonian (\ref{ham})  with $\mathbf{A} = \cosh{(\mathbf{\alpha S})}$ and $\mathbf{B} = \sinh{(\mathbf{\alpha T})}$ and $\alpha = 30$ ($\triangle$). Entropy computed for compact nonlocal hamiltonian (\ref{ham}), choosing the kinetic and pairing matrices by equations (\ref{NL_AB}); $\alpha = 5$ ($\diamond $); $\alpha = 30$ ($\Box$).  For $\alpha = 30$, the entropy shows a linear regime extending approximately to the locality scale and saturating due to finite size effects at larger scales.}
\end{figure}

We first address the issue of the different possible forms of nonlocality for fermions.  As a baseline, the entanglement entropy for a 1-d system described by the local pairing hamiltonian (\ref{ham}) is computed in the gapped phase with kinetic and pairing matrices, respectively, given $\mathbf{A} = \mathbf{S}$ and $\mathbf{B} = \mathbf{T}$. As expected, the entropy $S \sim O(1)$ as shown for a 100 site lattice in figure \ref{fig1}. Next we consider a model with nonlocality similar to the model introduced by Shiba and Takayanagi \cite{Shiba:2013jja} (our equation (\ref{NL_bosons})), but possessing the relativistic dispersion characteristic of a superconductor.  In hamiltonian (\ref{ham}) we choose $\mathbf{A} = \cosh{(\mathbf{\alpha S})}$ and $\mathbf{B} = \sinh{(\mathbf{\alpha T})}$.  The entropy (figure \ref{fig1}) is seen to saturate at a scale much shorter than the locality scale $\alpha = 30$. Finally we compute the entropy for a model with compact nonlocality given by hamiltonian (\ref{ham}) and choosing the kinetic and pairing matrices by equations (\ref{NL_AB}).  For the case of $\alpha = 30$, the entropy shows a linear regime extending approximately to the locality scale and saturating due to finite size effects at larger scales. Looking at $\alpha = 5$ (and $\alpha = 15$) it appears that the crossover to saturated behavior occurs at $L \approx \alpha$.  

Our conclusion is that nonlocality of a noncompact form in a fermionic hamiltonian with relativistic dispersion does not produce an entropy volume law, despite its similarities to relativistic bosonic models.  However, compact nonlocality appears to produce the desired entropy volume law.  

\begin{figure}[ht]
\includegraphics[width=7.5cm]{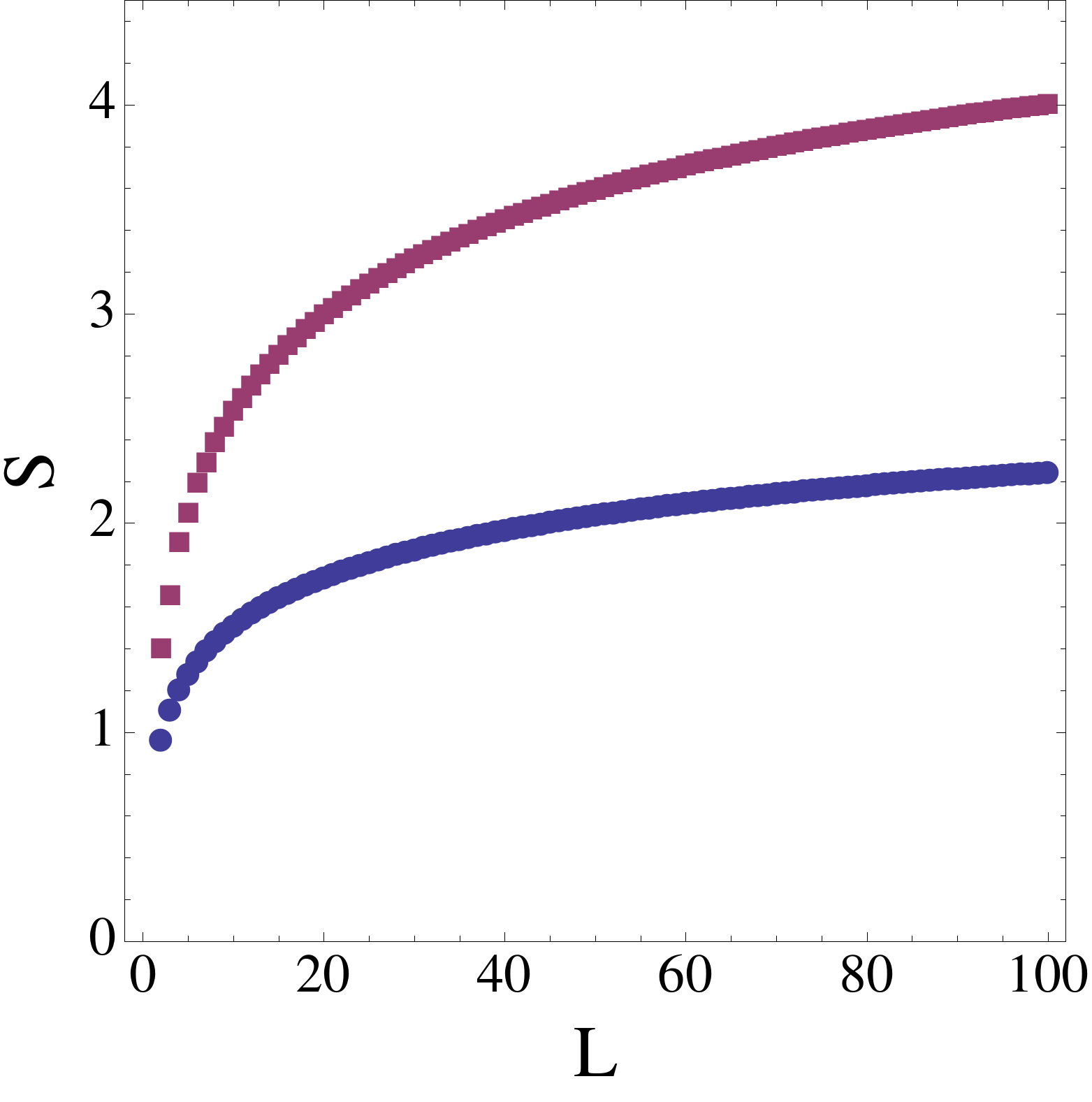}
\caption{\label{fig2} Entanglement entropy, $S$, of size $L$ subregion computed for the local fermion hamiltonian (\ref{local_ham}) on a 400 site 1-d lattice with an occupancy fraction $f = $ \#particles/\#sites $= 1/2$ ($\circ$). The effective central charge is computed to be $\ceff = 0.978 \approx 1$. Corresponding computation for nonlocal hamiltonian (\ref{NL_cos}) with $\alpha = 0.01$   ($\Box$).}
\end{figure}

\begin{figure}[ht]
\includegraphics[width=7.5cm]{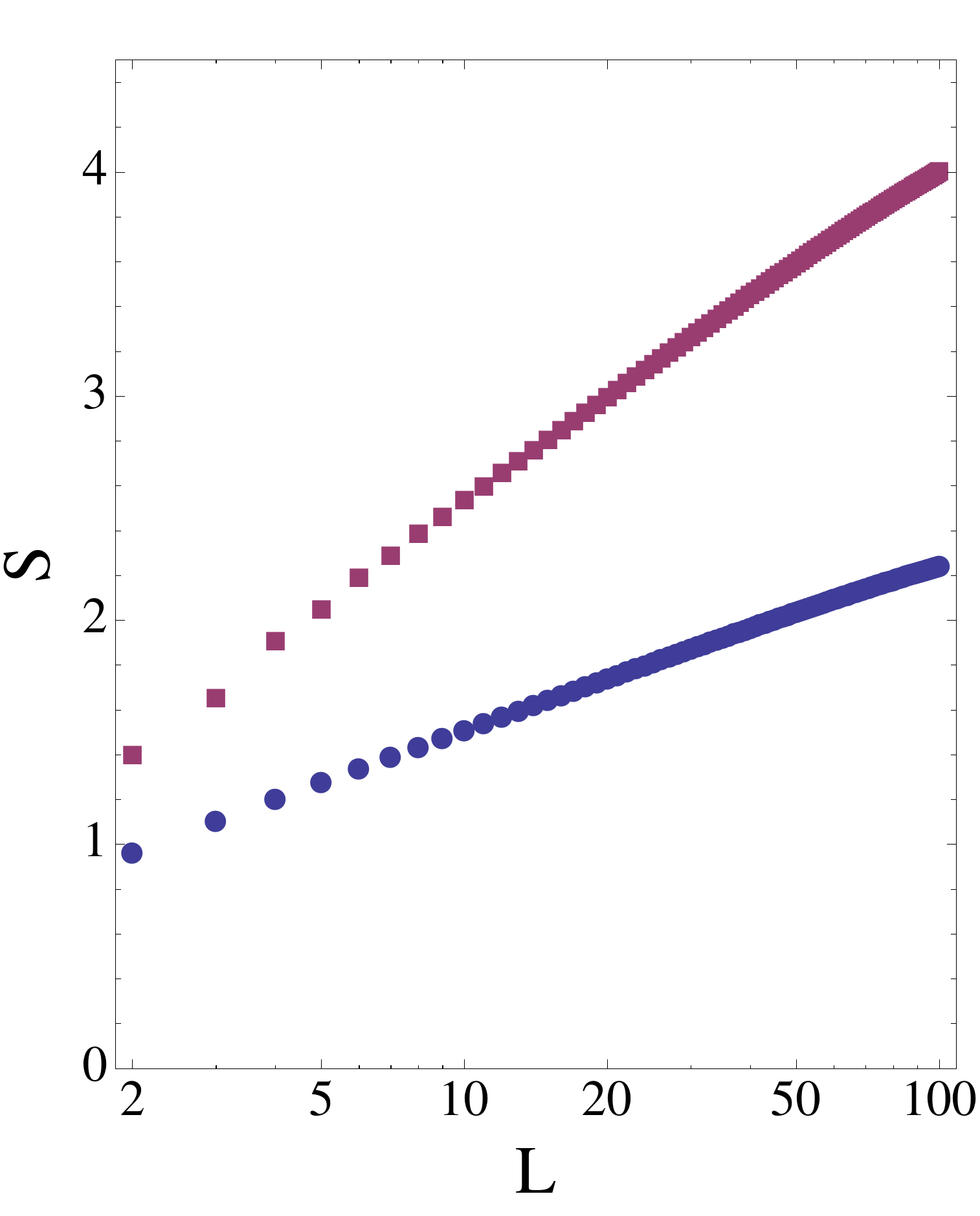}
\caption{\label{fig3} Log-linear plot of data from figure \ref{fig2}.}
\end{figure}

Because a relativistic dispersion, analogous to bosons, does not appear to be a necessary feature for a volume entropy law, we now turn to the simpler variants of nonlocal fermions described by hamiltonians (\ref{NL_cos}) and (\ref{NL_sin}).  For reference, the entropy is computed for a one-dimensional local (nearest neighbor hopping) model of spinless lattice fermions given by the hamiltonian 
\be
\label{local_ham}
H=\sum{c_x^\dagger S_{xy} c_y}
\ee
Figures \ref{fig2} and \ref{fig3} show the characteristic  logarithmic behavior for the entropy of 1-d gapless fermions; for the lower curve, the entanglement entropy is found numerically to follow: $S = c_0 + (c_{\rm{eff}}/3)\log{L} $, where $c_0$ is a constant and $\ceff \approx 0.978$ close to the expected value of the central charge $c=1$. 

Next, we compute the entropy for hamiltonian (\ref{NL_cos}) in the local limit $\alpha \rightarrow 0$.  Even though (\ref{NL_cos}) appears to be the lattice limit of hamiltonian (\ref{NL_fermions})---i.e. the matrix $ i [\partial]_{xy} = T_{xy}$ is the lattice first derivative---it is well known that this model does not express a single chiral lattice fermion.  Specifically, it exhibits two fermi points at $k = 0, \pi$, and therefore has two quasi-independent fermion species leading to a $c=1$ theory.  Diagonalizing this model for small $\alpha$ is in effect diagonalizing  $H=\sum{c_x^\dagger [\mathbf{T}^2]_{xy} c_y}$ or equivalently, $H=\sum{c_x^\dagger [\mathbf{S}^2]_{xy} c_y}$.  Entropy as a function of subsystem length is computed for hamiltonian (\ref{NL_cos}) with $\alpha = 0.01$ and displayed in figures \ref{fig2} and \ref{fig3}. The behavior is logarithmic: $S \approx (c_{\rm{eff}}/3)\log{L} $ with $\ceff= 1.96 \approx 2$. 

%It is instructive to write out the real space operators...two species of left movers; two species of right movers.

% $\circ$   $\Box$  $\diamond$    $\triangle$  %

\begin{figure}[ht]
\includegraphics[width=7.5cm]{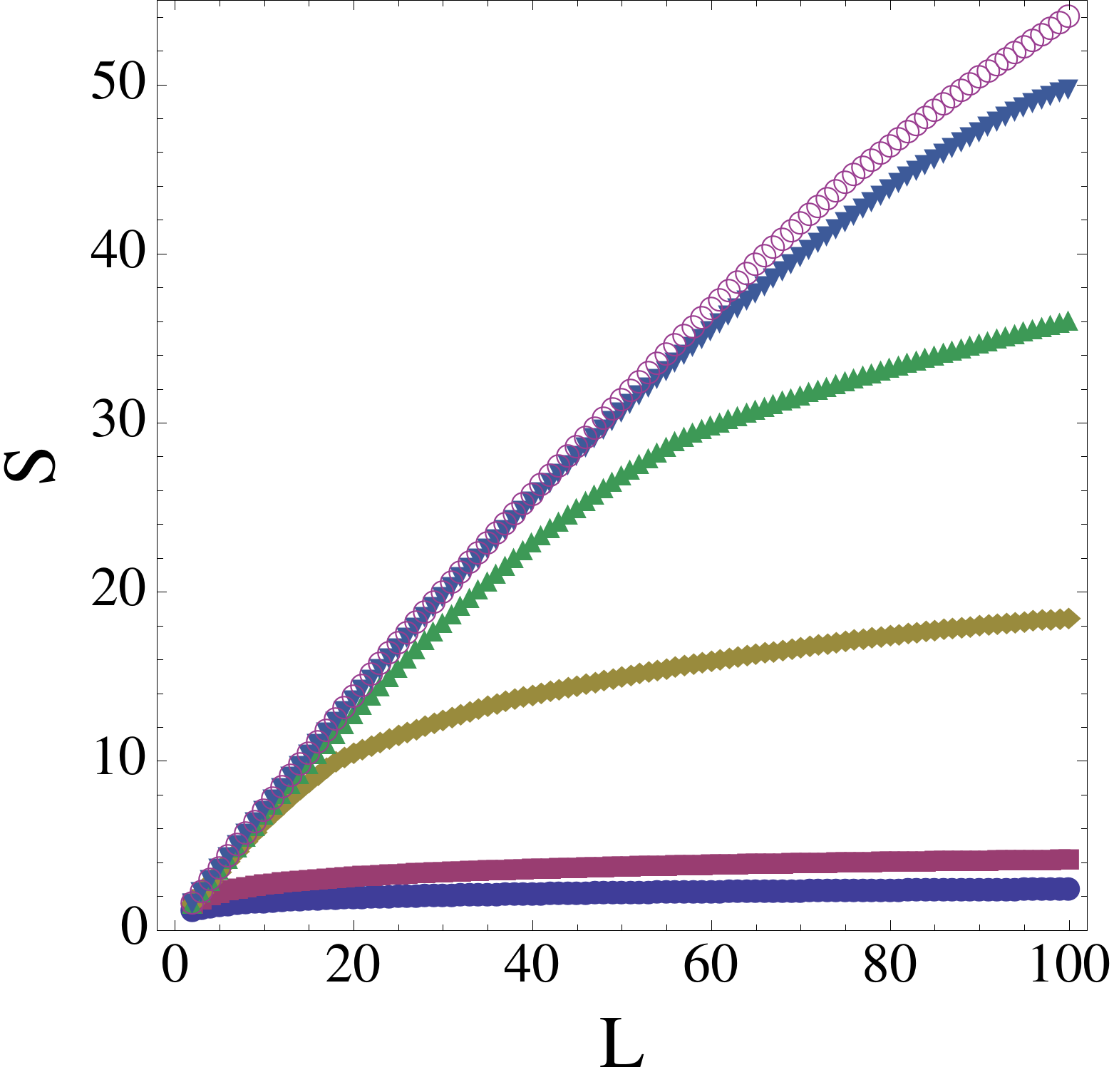}
\caption{\label{fig4}The dependence of the entropy upon the locality scale $\alpha$ in the compact nonlocal hamiltonian (\ref{NL_cos}). Entropy $S$ versus subregion size $L$ for local models from figure \ref{fig2} as a reference (solid $\circ$ and solid $\Box$). $S$ versus $L$ computed for hamiltonian (\ref{ham}) on a 400-site periodic 1-d lattice, with $\alpha = 10, 30, 50, 1400 $ ($\diamond$, $\triangle$, inverted $\triangle$, $\circ$ ).  For subregions much smaller than the locality scale ($L \ll \alpha$), approximately linear behavior is seen. }
\end{figure}

\begin{figure}[ht]
\includegraphics[width=7.5cm]{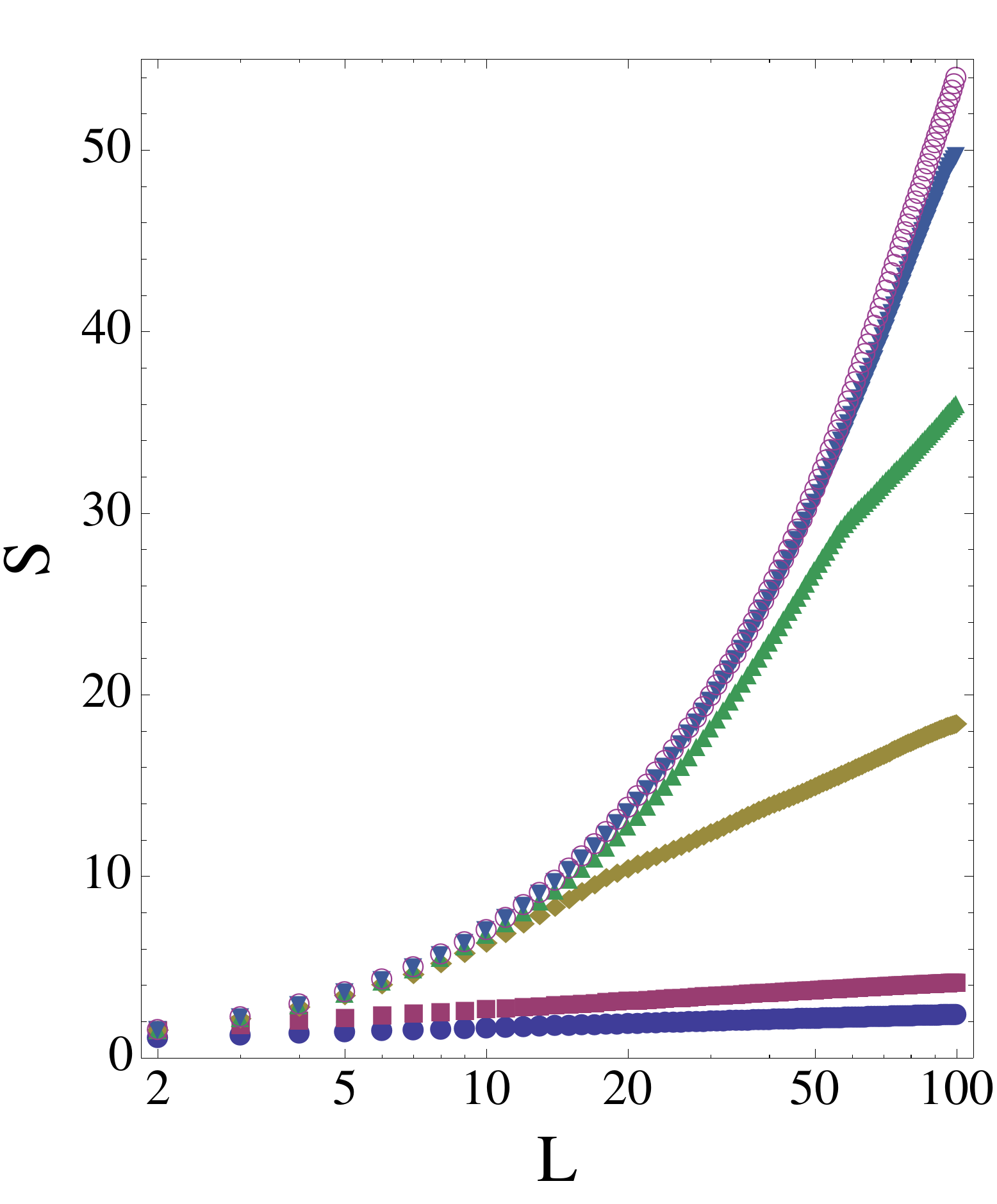}
\caption{\label{fig5}  Log-linear plot of data from figure \ref{fig4}.  For subregions much larger than the locality scale ($L >> \alpha$), entropy depends logarithmically upon $L$: $S \approx (\ceff/3) \log{L}$ where the effective central charge $\ceff \approx 1.17 \alpha$. }
\end{figure}

The dependence of the entropy upon the locality scale $\alpha$ in the nonlocal hamiltonian (\ref{NL_cos}) is studied in figure \ref{fig4}.  Looking at the large $\alpha$ cases first, there appears to be a nearly linear regime where the entropy behaves as $S \approx d L$ for $L \ll \alpha$ with $d \approx 0.5$. Analysis of the $L > \alpha$ regime is displayed on the log-linear plot of the same data in figure \ref{fig5}.  For this nonlocal model, there appears to be logarithmic regime, for $L > \alpha$, described approximately by $S \approx (\ceff/3) \log{L}$. The factor $\ceff$, which may associated with the number of fundamental degrees of freedom, is seen to be approximately the same as the locality parameter: $\ceff \approx 1.17 \alpha$.  This feature will be discussed in the next section.

\begin{figure}[ht]
\includegraphics[width=7.5cm]{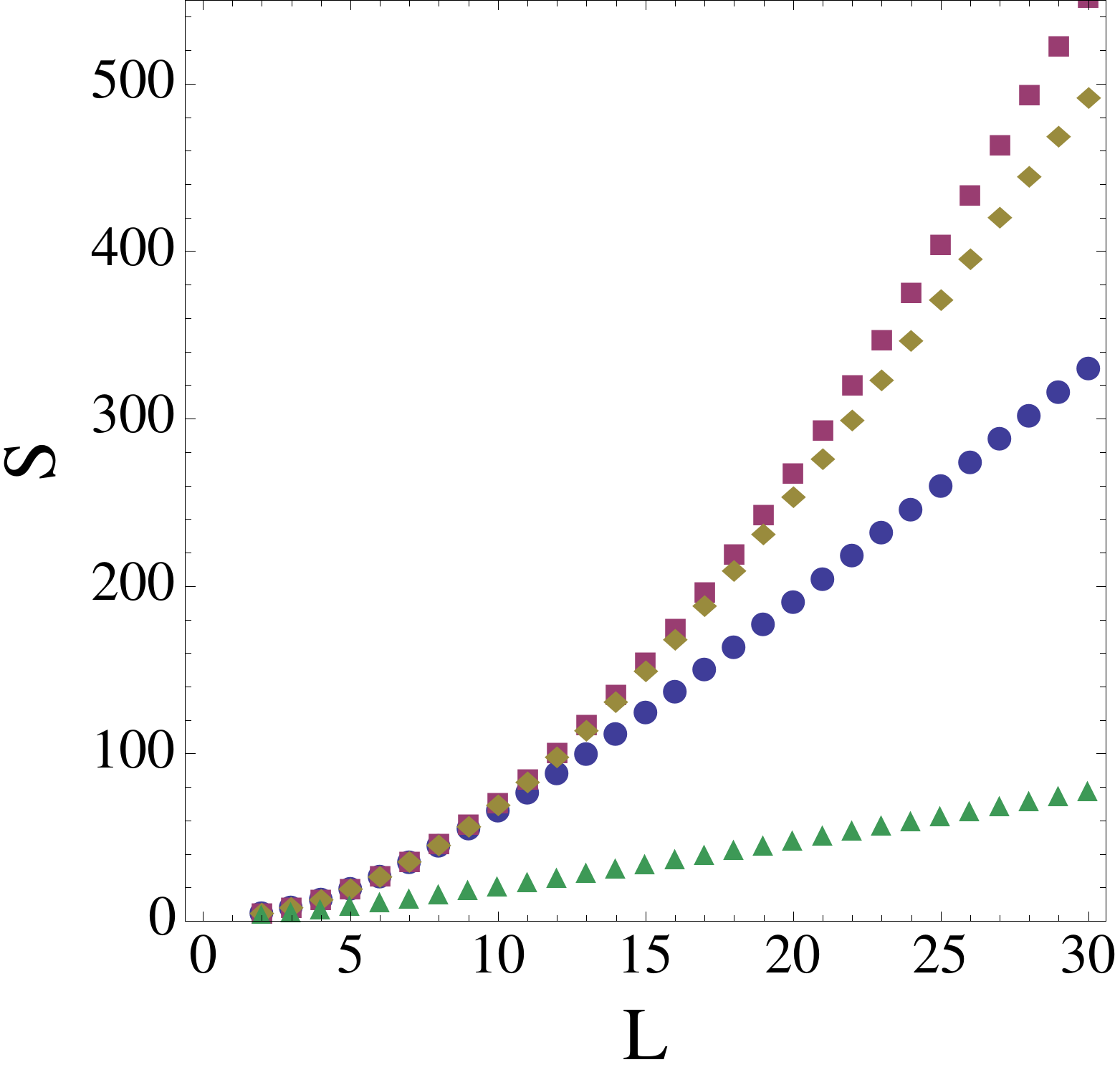}
\caption{\label{fig6}  The entanglement entropy computed for compact nonlocal model hamiltonian (\ref{NL_sin}) extended to 2-d as described in the text. Dependence of entropy, $S$, on linear size, $L$, of square subregion for a $61 \times 61$ site periodic lattice, computed for several locality parameters, $\alpha$. $S$ versus $L$ for local 2-d gapless fermions ($\triangle$), $\alpha = 5$ ($\circ$), $\alpha = 15$ ($\diamond$), $\alpha = 1400$ ($\Box$) .  The computations for large $\alpha$ show a $L \ll \alpha$ regime where the entropy appears to follow a volume law proportional to $L^2$, with a prefactor similar to the 1-d case: $S \approx dL^2 $ where $d\approx 0.5$.   }
\end{figure}

\begin{figure}[ht]
\includegraphics[width=7.5cm]{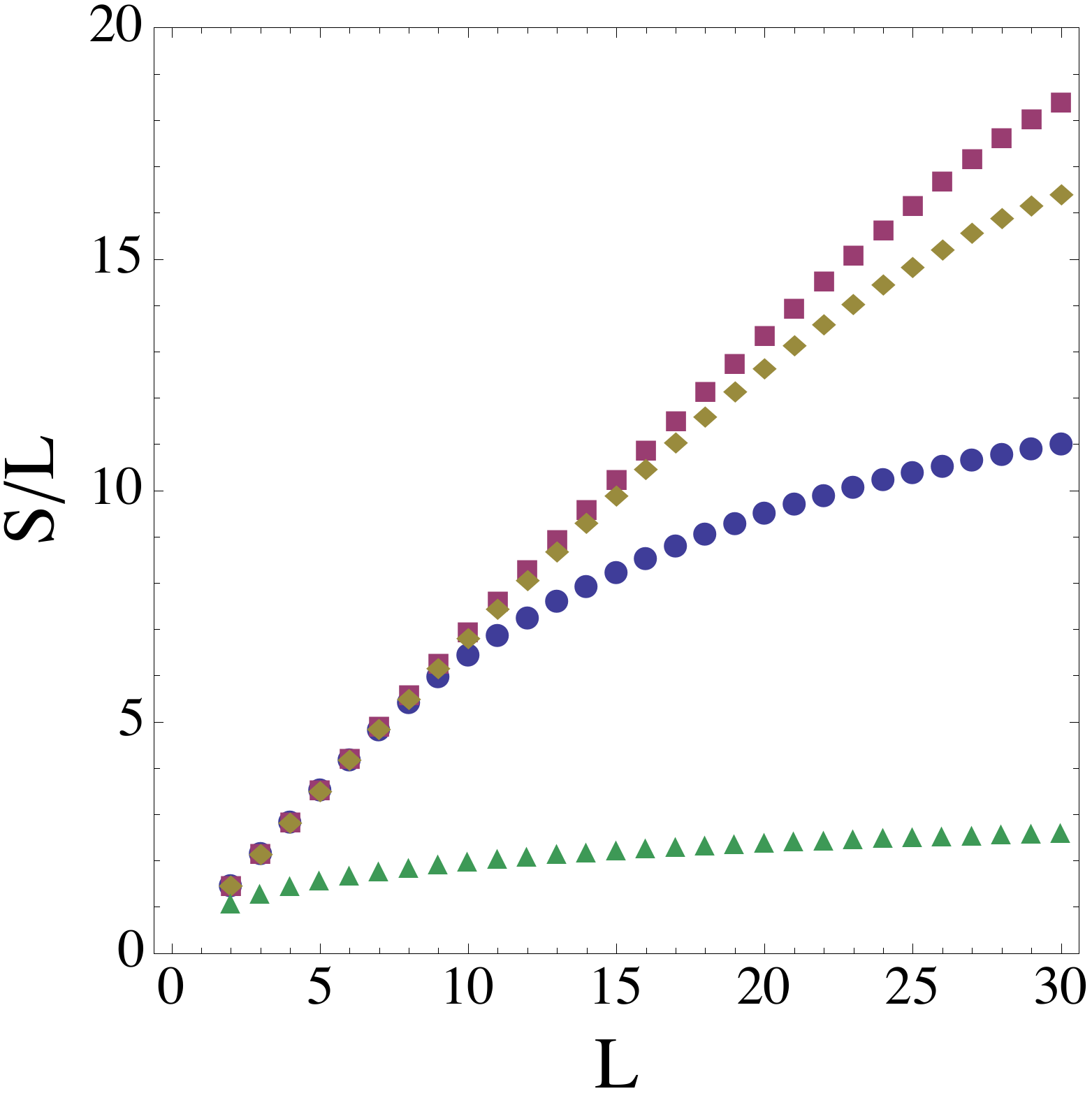}
\caption{\label{fig7} Data from figure \ref{fig6} rescaled by $L$.  For $L > \alpha$, $S/L$ clearly becomes sublinear but monotonically increasing toward saturation at $L=30$.}
\end{figure}

To extend these computations to 2-d square lattices, hamiltonian (\ref{NL_sin}) is used with the lattice differential operator $S_\mathbf{x,y} = \delta_{\<\mathbf{x,y}\>}$ where $\<\mathbf{x,y}\>$ denotes all possible nearest neighbor 2-d lattice vector points $\mathbf{x}$ and $\mathbf{y}$ .  The choice of the symmetric matrix, $\mathbf{S}$, avoids cross terms in the higher derivatives of the $\sin{\mathbf{\partial}}$ expansion, such as $\partial_{x_1}\partial_{x_2}$ where $x_1$ and $x_2$ are orthogonal lattice directions. With this choice, the discrete derivative operator in hamiltonian (\ref{NL_sin}) may be thought of as the lattice version of $\sin{\nabla^2}$
 
\begin{figure}[ht]
\includegraphics[width=7.5cm]{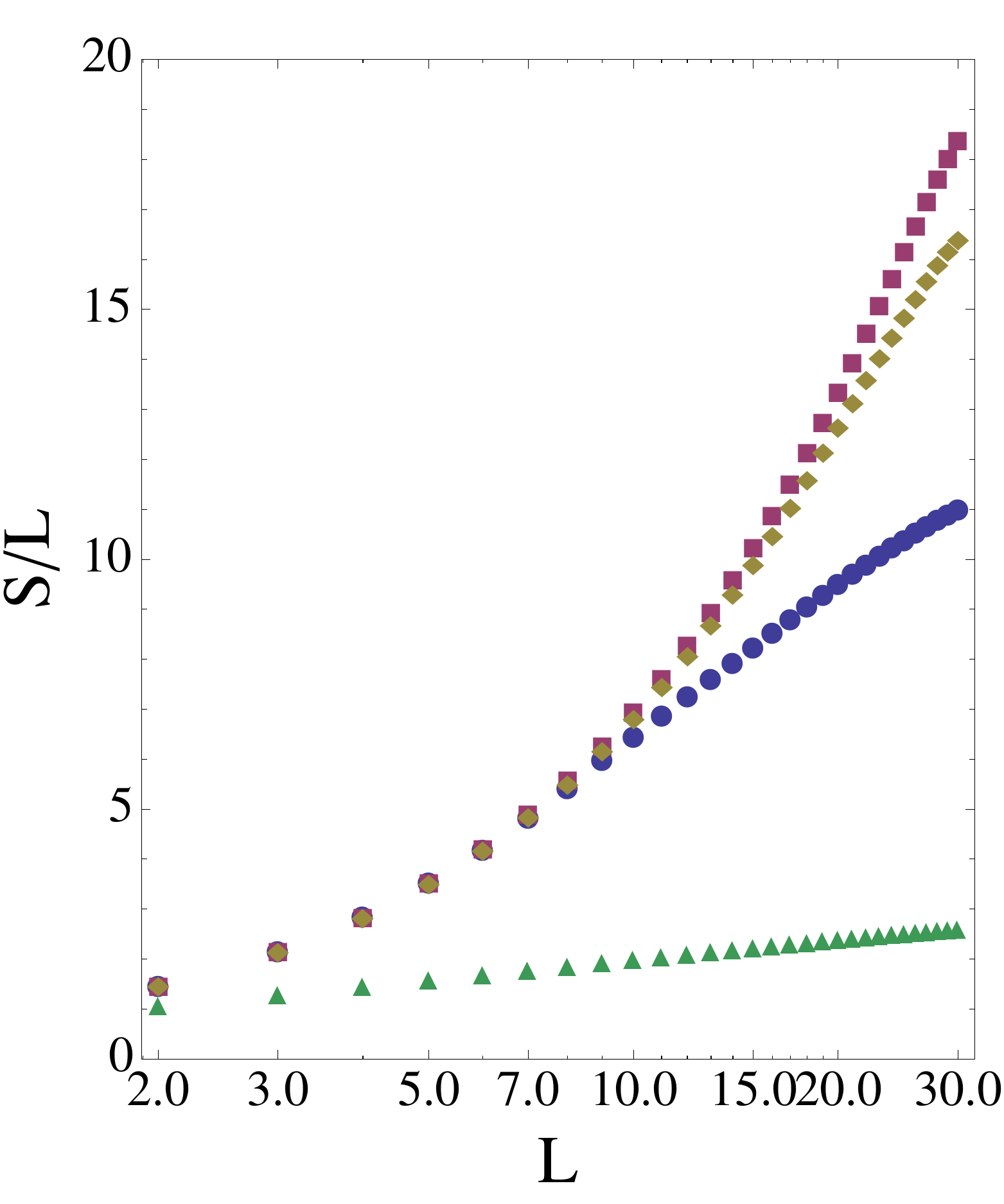}
\caption{\label{fig8}Data from figure \ref{fig7} on log-linear plot.}
\end{figure}

Using a $61 \times 61$ site periodic lattice, the entanglement entropy was computed for a series of square subregions of linear size $L$. Figure \ref{fig6} shows the results for several locality factors, $\alpha$.  The computations for large $\alpha$ show a $L << \alpha$ regime where the entropy appears to follow a volume law proportional to $L^2$, with a prefactor similar to the 1-d case: $S \approx dL^2 $ where $d\approx 0.5$.  The behavior in the vicinity of $L \sim \alpha$ is best seen in figure \ref{fig7}; for $L > \alpha$, $S/L$ clearly becomes sub-linear.  Examining $S/L$ on a log-linear plot (figure \ref{fig8}) suggests that $S \propto L \log{L}$ for $L > \alpha$, a crossover to the anomalous fermion area law \cite{GK2006}. To analyze this behavior, we note that the entropy of local model is known to follow the following exact analytic form: $S = (2/3)\ceff L \log{L}$ \cite{Haas2006,BCS2006}, where $\ceff=1$ anticipating the relationship of the 2-d anomalous area law to the underlying 1-d tomographic fermions \cite{levine_miller2008,bs2010}.  Computational limitations restrict our computations to sizes no bigger than $61 \times 61$, thus saturation of the entropy appears at subregions of size $L \approx 30$.  However, for locality factors between $\alpha = 5$ and $\alpha = 15$ there is enough data between before saturation to reliably fit $S/L$ logarithmically and determine: 
\be
S \approx \frac{2}{3} \ceff L \log{L}
\ee
where $\ceff \approx 1.26 \alpha$, similar to the relationship in 1-d: $\ceff= 1.17 \alpha$.  In the following section we will discuss the identification of $\alpha$ with the number of fundamental degrees of freedom and the relevance of these fits of 1-d and 2-d entropy data.

In summary, we find the entropy dependence upon linear subregion size ($L$) for the nonlocal 1-d hamiltonian (\ref{NL_cos}) to be: 
\bea
\label{law1d}
S &=& AL \,\,\,\,\,\,\,\, (L<\alpha) \\
S &=& A\alpha + B \frac{\alpha}{3} \log{L} \,\,\,\,\,\,\,\, (L>\alpha) \nonumber
\eea
and for the nonlocal 2-d hamiltonian (\ref{NL_sin}): 
\bea
\label{law2d}
S &=& AL^2 \,\,\,\,\,\,\,\, (L<\alpha) \\
S &=& A\alpha^2 + B^\prime \frac{2\alpha}{3} L\log{L} \,\,\,\,\,\,\,\, (L>\alpha) \nonumber
\eea
where $A \approx 0.5$, $B \approx 1.17$ and $B^\prime \approx 1.26$.

\section{Discussion}
We would like to understand the features of compact nonlocal models that have appeared so far:  (1) emergence of the locality parameter, $\alpha$, as a measure of the number of fundamental degrees or effective central charge, $\ceff \approx \alpha$ (2) a conventional fermionic area law for regions $L \gg \alpha$ and (3) the volume entropy law for length scales shorter the locality length ($L \ll \alpha$). 

Consider hamiltonian (\ref{NL_fermions}) on a periodic interval $[0,R]$ (and setting the energy scale, $\epsilon=1$):
\be
H = \sum_{xy} {c^\dagger_x [\cos{\partial}]_{xy} c_y}  
\ee
The energy spectrum of the model is given by $E_n = \cos{(\alpha \sin{k_n})}$, where $k_n = 2 n \pi/R$ and $n \in [-R/2,R/2]$. This model has low energy single particle states with linear dispersion in the vicinity of $\alpha \sin{\frac{2 \pi n}{R}} \approx \frac{2m+1}{2}\pi$ where $m = 0,\pm 1,\pm 2\ldots$. Much like the conventional (local) 1-d spinless fermion model, these low energy sectors are independent and each contribute an additive  entropy that is logarithmic in the subsystem size.  Since $n$ itself is bounded by $\pm R/2$, the number of low energy domains may be estimated to be: $m_{\rm max} \approx  4\alpha/\pi$.  The low energy points of $E_n$ come in pairs, dispersing with positive and negative velocity; assigning a central charge $c=1/2$ to each point give an effective central charge for the nonlocal model $\ceff \approx 2\alpha/\pi$, similar to our numerical estimates from figure \ref{fig5}.  Regarding $m$ as a "flavor" index, low energy excitations at a momentum $k_m$, given implicitly by $\sin{k_m} \approx \frac{m \pi}{\alpha}$, define a wavepacket of size $\xi \gg \alpha$. To create wavepackets with $\xi \ll \alpha$ involves mixing momentum states over a range $O(1/\alpha)$, that is, mixing states over two or more adjacent flavor indices. Thus we expect the large $L$ logarithmic behavior to cross over to some other behavior when $L \approx \alpha$.

In the bosonic nonlocal models proposed by Li and Takayanagi \cite{Li:2010dr}, the entropy was computed by a replica/orbifold technique or by directly investigating the structure of the exact reduced density matrix \cite{Shiba:2013jja}. For the model of compact nonlocal fermions presented here, we have not been able to compute the entropy analytically.  However, we are able to infer the volume law entropy from properties of the single particle correlation function in the limit of strong nonlocality: $\alpha \gg R,L$. 

Consider the lattice correlation function for $\alpha \gg R$:
\be
G_{xy} = \frac{1}{R}\sum_{m=0}^{R-1}{\< c^\dagger_m c_m \> e^{2\pi i m(x-y)/R}}
\ee 
where the Fermi distribution $\< c^\dagger_m c_m \>$ may be regarded as a random binary string with variance $\sigma_0^2 = 1/4$. The discrete fourier transform of a random function (normalized as above) is another random function with rescaled variance $\sigma^2 = \sigma_0^2/(2R)$. Computation of the entanglement entropy in a subregion depends upon the eigenvalues of the correlation function restricted to the subregion. The correlation function is a Toeplitz matrix with gaussian distributed random entries.  Recently it was proven that such matrices have gaussian distributed random eigenvalues \cite{random_Toeplitz}. The von Neumann entropy of $L$ such eigenvalues is proportional to $L$ thus establishing extensive entropy.

\begin{figure}[ht]
\includegraphics[width=9.5cm]{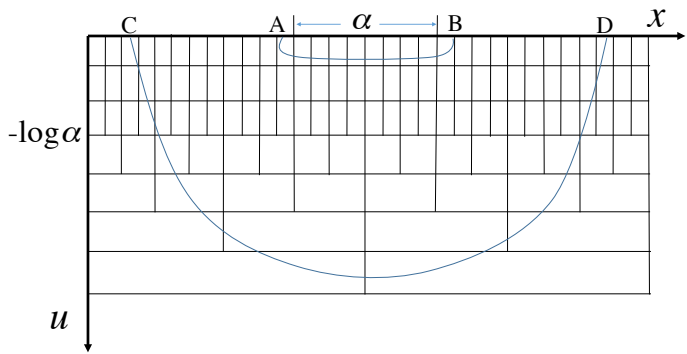}
\caption{\label{fig9} Cartoon of AdS-flat space hybrid based upon a UV extension of metric (\ref{AdS_metric}) to flat space. The entanglement of regions comparable to the locality scale ($L \approx \alpha$), such as $L_{\rm AB}$ depicted above, involve geodesics through flat space with minimal extension into the bulk. Regions larger than the locality scale ($L \gg \alpha$), such as $L_{\rm CD}$ depicted above, have geodesics that penetrate into the AdS region. }
\end{figure}

\begin{figure}[ht]
\includegraphics[width=9.5cm]{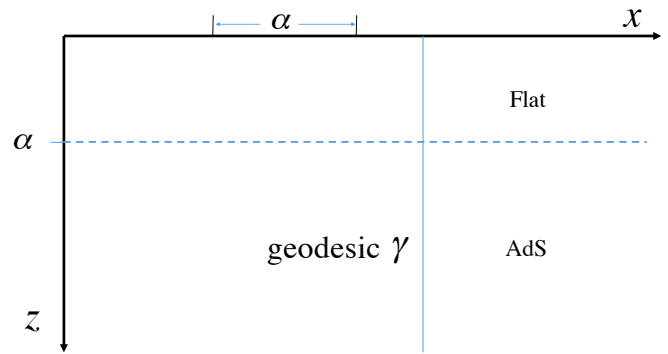}
\caption{\label{fig10} Cartoon of AdS-flat space hybrid in Poincare coordinates suggesting the metric (\ref{NL_metric}). The size, in $z$, of the flat space region must be proportional to the locality scale $\alpha$ of the boundary theory, as well as the AdS radius $\kappa \approx \alpha$ to produce the correct entropy-length behavior (\ref{law1d}) in the ground state of nonlocal hamiltonian (\ref{NL_cos}). }
\end{figure}

%Modular hamiltonian and geodesics.

 \begin{figure}[ht]
\includegraphics[width=9.0cm]{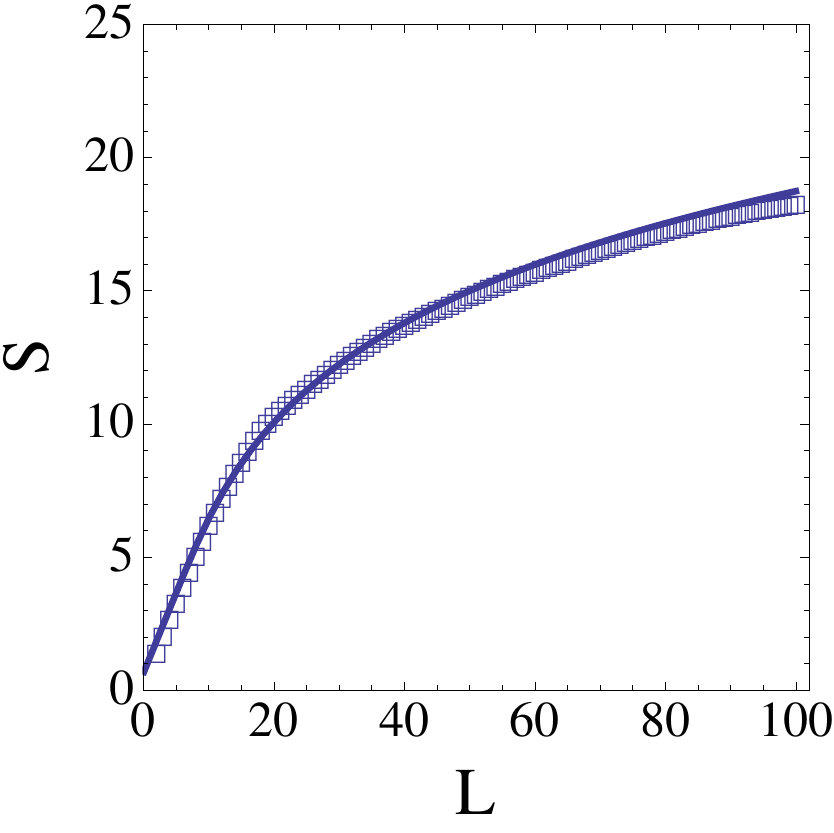}
\caption{\label{fig11} Numerical integration of length along geodesic $\gamma$ using metric (\ref{NL_metric}) compared with numerical computation of entropy of figure \ref{fig4}, $\alpha = 10$. Following equation (\ref{S_comp}), $S = a\int_0^L{dz\sqrt{\tanh{\frac{\alpha_{\rm c}^2}{z^2}}}} + b$ where $\alpha_{\rm c} = 9$, $a = 0.6$ and $b = 0.7 $}
\end{figure}

Within the holographic picture of Ryu and Takayanagi \cite{RT2006}, our central result (equation \ref{law1d}) suggests a bulk space with a crossover between flat space in the UV and AdS space in the IR to accommodate the crossover in entropy between extensive and logarithmic behaviors as the length of the subregion increases. By the Brown-Henneaux relation, the radius of AdS$_3$, $\kappa$, is proportional to the central charge of the boundary theory, $c = \frac{3\kappa}{2G}$, where $G$ is the 3-d Newton constant. But the effective central charge in our model of compact nonlocality is {\sl also} proportional to the locality scale, $\alpha$, and the bulk metric must capture this property.  Figure \ref{fig9}  is a cartoon of the spatial slice of such a space based upon the structure of the pure AdS$_3$ metric
\be
\label{AdS_metric}
ds^2 = \kappa^2(du^2 + \frac{e^{2u}}{\epsilon^2}dx^2)
\ee
showing the space modified by a flat space slice of width $\alpha$ at the boundary. Thus the entanglement of regions shorter than the locality scale ($L \ll \alpha$) involve geodesics through flat space with minimal extension into the bulk, whereas regions larger than the locality scale ($L \gg \alpha$) have geodesics that penetrate into the AdS region. 

Looking at the behavior of the 1-d entropy (\ref{law1d}) for $L$ larger than the locality scale $\alpha$, it is seen that $\alpha$ enters the expression in two additive roles: the saturation of the extensive entropy at the locality scale $L \approx \alpha$ and the dependence of the effective central charge on $\alpha$ in the logarithmic term. Switching to Poincare coordinates in figure \ref{fig10}, consider the vertical portion of the geodesic $\gamma$, appropriate for computing the large $L$ entropy ($L \gg \alpha$) by the Ryu-Takayanagi construction:  
\be
\label{RT}
S = \frac{1}{2G}\int_\gamma{ds}
\ee
A critical feature of this geometry is then that the size (in $z$) of the flat region must be proportional to the radius of the asymptotic AdS region to reproduce both terms of equation (\ref{law1d}) for the entropy: $S = A\alpha + B(\alpha/3)\log{L}$.  To this end we propose the following spatial slice metric (in Poincare coordinates, $z = \epsilon e^{-u}$) capturing these features,
\be
\label{NL_metric}
ds^2 = \tanh{(\frac{\alpha_{\rm c}^2}{z^2})}(dx^2 + dz^2)
\ee
where we have replaced the AdS radius with the locality scale, $\alpha_{\rm c}$. Note that $\alpha_{\rm c}$ in this expression has dimensions of length corresponding to the continuum hamiltonian (\ref{NL_fermions}), and in contrast to the dimensionless $\alpha$ appearing in our lattice computations. 

Studying (\ref{NL_metric}) for constant $x$, there is the small $z$ regime ($0 < z < \alpha_{\rm c}$) in which $ds \sim dz$, and a large $z$ regime where $ds \sim (\alpha_{\rm c}/z) dz$ as $z \rightarrow \infty$.  If we consider a large $L$ subregion, the entropy may be approximated by integrating the Ryu-Takayanagi relation along the constant $x$ geodesic $\gamma$ pictured in figure \ref{fig10}, and imposing an IR cut off of $L$. 
\bea
\label{S_comp}
S &=& \frac{1}{2G} \int_0^L{dz\sqrt{\tanh{\frac{\alpha_{\rm c}^2}{z^2}}}} \\
&\approx& \frac{\alpha_{\rm c}}{2G} + \frac{\alpha_{\rm c}}{2G} \log{\frac{L}{\alpha_{\rm c}}}
\eea
in reasonable agreement with our results for the 1-d entropy (\ref{law1d}), making the identification $\ceff = \alpha = \frac{3 \alpha_{\rm c}}{2G}$. The integral appearing in (\ref{S_comp}) cannot be done analytically (there is an essential singularity at $z=0$) and we resort to a numerical evaluation. Figure \ref{fig11} provides a comparison of a numerical integration of the geodesic length appearing in equation (\ref{S_comp}) with our numerical lattice computations for a representative intermediate $\alpha$ case. There is reasonable agreement, but it should be noted that the geodesic $\gamma$ is only an approximation of the putative geodesic corresponding to the finite $L$ lattice computation.  Until further studies, we conclude that the proposed metric is a qualitative description of the gravitational dual to the nonlocal model (\ref{NL_cos}).

%The metric (\ref{NL_metric}) contains an essential singularity at $z=0$. 

%It is interesting to note (BB) that the geometry of extremal black branes underlying the original AdS/CFT construction have a feature resembling this crossover between flat space and AdS space, however entropy computed for those metrics adapted to AdS$_3$ do not appear to match our results. 

\section{Conclusion and Further Directions}

In this manuscript we have suggested nonlocality for fermions based upon a hamiltonians (\ref{NL_cos}) and (\ref{NL_sin}) incorporating periodic functions of the momentum operator. These models exhibit a crossover between volume and anomalous (logarithmic) area law behavior at scales larger than the locality scale, $\alpha$. The locality scale plays two roles. Firstly, it is the scale at which the volume law saturates. But owing to the periodic nature of the hamiltonian it is also a measure of the number of fundamental degrees of freedom in the theory---the effective central charge, $\ceff$.

In 1-d we suggest that the hamiltonian (\ref{NL_cos}) has a geometric interpretation within the Ryu-Takayanagi holographic picture of a hybrid AdS space in the IR and Euclidean space in the UV.  Because of its dual role, the locality scale determines both the size of the UV flat slice and the radius of the asymptotic AdS space.  We propose a specific metric in 3-d bulk space that captures the entropy/length relation for the ground state of this nonlocal model.

There are several obvious directions for future work. Clearly, it would  be desirable to compute the entropy analytically for the continuum version of our nonlocal model (\ref{NL_fermions})  on an orbifold, parallel to the calculation of \cite{Li:2010dr} for a nonlocal scalar QFT. For our model, note that a bosonized action only corresponds to set of ($\ceff$) linearized fermion modes; the nonlocal, extensive, behavior comes from the part of the dispersion {\sl between} the Fermi points. Thus a replica/orbifold computation involves computing the spectrum of the nonlocal Dirac type operator $\cos{(i \alpha \partial_x)}$. It may also be interesting to investigate whether a compact dispersion in a scalar boson model analogous to \cite{Li:2010dr} also exhibits a volume law, and whether the locality scale also proportional to the effective central charge.

The combination $i \alpha \partial_x$ in the nonlocal hamiltonian (\ref{NL_fermions}) also invites the following anisotropic generalization of in $d$ dimensions,
\be
\label{NL_HD_fermions}
H =  \epsilon\int{d^dx \psi^\dagger(x) \cos{(i\alpha^\mu \partial_\mu)} \psi(x)}
\ee
Anisotropy comes at the cost of destroying the nonrelativistic, quadratic dispersion limit as $\alpha \rightarrow 0$ in dimensions $d>1$. However it has been pointed out \cite{BB} that if the set  $\{\alpha^\mu \}$ are extended to form a Clifford algebra $\{\alpha^\mu,\alpha^\nu \} = \delta_{\mu \nu}$, the nonrelativistic quadratic limit is maintained.  Thus an interesting outcome is that nonrelativistic lattice fermions may be realized as the $\alpha \rightarrow 0$ limit of nonlocal fermions with additional isospin degrees of freedom. To include interactions, one possibility is to exploit the well defined (random) UV behavior of the propagators when $\alpha \rightarrow \infty$  and develop a perturbation expansion in $1/\alpha$.  With interactions, an inhomogeneous locality parameter $\alpha = \alpha(x)$ might also be a suitable variational parameter. 

Lastly, there are several, related, nonequilibrium properties that might be explored. In a non-translationally invariant model, the thermalization time scale \cite{Magan:2015yoa} and closely related scrambling time scale \cite{Hayden:2007cs, Magan:2016ojb} have been computationally studied for nonlocal fermions. It seems likely that subsystems smaller than the locality scale in our model would satisfy the Eigenstate Thermalization Hypothesis (ETH), as do globally nonlocal fermions in the cited studies, because the UV behavior of our correlation functions is effectively disordered. For this reason, it would be natural to study the evolution of entanglement in a variety of nonequilibrium initial conditions \cite{cardy_quench, Igloi_disorder, dis_dynamics, stephan_quench, Peschel_rev2009, FCS_disorder} 

\section*{Acknowledgement} GL wishes to thank Benjamin Burrington and Adam Durst for many interesting discussions.

%In the continuum version of the compact nonlocal model (equation \ref{NL_fermions}), we expect the dimensionful locality scale $\alpha \approx \kappa$, in that the Planck scale, $G$, would play the role of a lattice cut off. 
%replica  continuum model - Perhaps noncompact form works?
%dirac op in curved space; How is AdS (-cosmo constant) stabilized? compute the source from metric. 

\end{document}